\documentclass[prd,aps,showpacs,eqsecnum,floatfix,nofootinbib,preprint,tightenlines]{revtex4}

\usepackage{latexsym}
\usepackage{epsf}
\usepackage{graphicx}
\usepackage{slashed}
\usepackage{amsmath}
\usepackage{amssymb}
\usepackage{multirow}

\def\cbo{{\,\raise-.15ex\Sc [\,}}                       

\def\ddt#1{{\buildrel {\hbox{\LARGE .\kern-2pt.}} \over {#1}}}

\newcommand{\gA}{g_A}
\newcommand{\gAb}{\bar{g}_A}
\newcommand{\tm}{t_{\rm m}}

\long\def\symbolfootnote[#1]#2{\begingroup%
\def\thefootnote{\fnsymbol{footnote}}\footnote[#1]{#2}\endgroup}

\long \def \blockcomment #1\endcomment{}

\def\vp{{\vec p}}

\newcommand{\pref}[1]{(\ref{#1})}

\newcommand{\mN}{M_N}

\begin{document}
\hyphenation{fer-mio-nic per-tur-ba-tive pa-ra-me-tri-za-tion
pa-ra-me-tri-zed a-nom-al-ous}

\renewcommand{\thefootnote}{$*$}

\preprint{YITP-16-81}

\title{Nucleon-pion-state contribution in lattice calculations of the nucleon charges $g_A,g_T$ and $g_S$}

\author{Oliver B\"ar$^{a}$} 
\affiliation{$^a$Yukawa Institute for Theoretical Physics, Kyoto University,
\\Kitashirakawa Oiwakechou, Sakyo-ku, Kyoto 606-8502, Japan\\
}

\begin{abstract}
\vspace{0.5cm}

We employ leading order covariant chiral perturbation theory to compute the nucleon-pion-state contribution to the 3-point correlation functions one typically measures in lattice QCD to extract the isovector nucleon charges $g_A,g_T$ and $g_S$. We estimate  the impact of the nucleon-pion-state contribution on both the plateau and the summation method for lattice simulations with physical pion masses. The nucleon-pion-state contribution results in an overestimation of all charges with both methods. The overestimation is roughly equal for the axial and the tensor charge, and about fifty percent  larger for the scalar charge. 

\end{abstract}

\maketitle

\renewcommand{\thefootnote}{\arabic{footnote}} \setcounter{footnote}{0}

\newpage
\section{Introduction}\label{Intro}

Lattice QCD has made enormous progress in the last years  due to computational advances and algorithmic improvements \cite{Schaefer:2012tq}. This has led many lattice QCD collaborations to pursue numerical lattice simulations with pion masses close to or at their physical values \cite{Abdel-Rehim:2015owa,Bazavov:2014wgs,Durr:2013goa,Aoki:2009ix}. Such ``physical point simulations'' require no or only a short chiral extrapolation, so uncertainties associated with this step are essentially eliminated. This benefit is worth the high numerical costs  these kind of simulations involve.

As advantageous as physical point simulations are, some complications get more severe the smaller the pion masses are. The signal-to-noise problem \cite{Lepage:1991ui} gets worse and prevents large euclidean time separations in many correlation functions. In addition, the smaller the pion mass the more pronounced is the contamination due to multi-particle-states in correlation functions one measures in lattice simulations. For example, the lattice simulations carried out so far strongly suggest that many nucleon structure observables suffer severely from excited-state contaminations.\footnote{See the recent reviews \cite{Syritsyn:2014saa,Green:2014vxa,Constantinou:2015agp} and references therein.}
The associated systematic uncertainty may significantly compromise the huge numerical effort that goes into physical point simulations. 

It has been pointed out in Refs.\ \cite{Tiburzi:2009zp,Bar:2012ce} that chiral perturbation theory (ChPT) can be employed to compute multi-particle-state contributions involving light pions. Following up Refs.\ \cite{Bar:2015zwa,Tiburzi:2015tta} we apply this idea here to the nucleon 3-point (pt) functions used to measure the non-singlet axial, tensor and scalar charge of the nucleon. We compute the nucleon-pion-state ($N\pi$) contributions to these observables in covariant ChPT to leading order (LO) in the chiral expansion.    
The low-energy-coefficients (LECs) entering at this order are known very well from phenomenology, so we obtain definite results for the $N\pi$ contribution to all three charges, estimated either by the plateau or by the summation method. Even if higher order corrections  and contributions from resonances will be substantial we do obtain quantitative estimates for the impact of the nucleon-pion states  on the determination of the various nucleon charges.  

\section{Nucleon 3-pt correlators in ChPT}

\subsection{Basic definitions}
In the following we consider QCD with degenerate quark masses for the light up and down quark. The spatial volume is assumed to be finite with spatial extent $L$ and periodic boundary conditions are imposed. We work in euclidean space time and the time extent is taken infinite.

We are interested in the 3-pt functions 
\begin{equation}\label{Def3pt}
G_{{\rm 3pt},X}(t,t') = \int d^3x\int d^3y \,\Gamma'_{X,\alpha\beta} \langle N_{\beta}(\vec{x},t) O_{X}(\vec{y},t') \overline{N}_{\alpha}(\vec{0},0)\rangle\,.
\end{equation}
Here $N, \overline{N}$ are interpolating fields for the nucleon and $\Gamma'_{X}$ denotes a spin projection matrix specified below. $O_X$ denotes the vector current ($X=V$), the axial vector current ($A$), the tensor ($T$) or the scalar density ($S$). We consider the flavor non-singlet case only, so $O_X$ carries an open flavor index which is suppressed in \pref{Def3pt}. More precisely, we choose the nucleon to be the proton which implies
\begin{eqnarray}
O_{X} = \overline{q}\,\Gamma_{\!X}\sigma^3 q,
\end{eqnarray}
with the quark doublets $q=(u,d)^T$, $\overline{q} = (\overline{u},\overline{d})$, the third Pauli matrix $\sigma^3$, and the usual gamma matrix combinations $\Gamma_X = \gamma_{\mu}, \gamma_{\mu}\gamma_5, \sigma_{\mu\nu},1$ for $X=V,A,T,S$. In the following we will be interested only in the spatial components in case of the axial vector current and the tensor, and in the zero component in case of the vector current. The projection matrices $\Gamma'_{X}$ for these operators are collected in table \ref{table:projectors}. For the axial vector current and the tensor we consider the averaged correlation function where the average is taken over the spatial components.\footnote{The results in section \ref{ssect3ptfunctions} assume a slightly simpler form for the averaged correlator than for the one with fixed spatial components. However, the final results for the nucleon-pion-state contribution are the same in both cases.}
\begin{table}[tp]
\begin{center}
\begin{tabular}{c||c|c|c|c}
$X$ & $\,V_0\,$& $A_k$ &$T_{kl}$ &$\,S\,$  \\\hline
$\Gamma'_X$ &$ \Gamma$ &$\Gamma \gamma_{k}\gamma_5$& $i\epsilon_{klm}\Gamma \gamma_5\gamma_m$& $\Gamma$
\end{tabular}
\end{center}
\caption{The projection matrices $\Gamma'_X$ entering the definition \pref{Def3pt}, with $ \Gamma =(1+\gamma_0)/2 $ and $k,l,m=1,2,3$ (spatial indices only).}
\label{table:projectors}
\end{table}
Since we consider equal up and down type quark masses the vector current is conserved. Thus, the 3-pt function involving its zero component is simply the 2-pt function times the conserved charge and not very interesting. However, charge conservation provides a non-trivial check on the calculation in section \ref{ssect3ptfunctions}.

In addition to \pref{Def3pt} we will also need the 2-pt function  
\begin{equation}\label{Def2pt}
G_{\rm 2pt}(t) = \int d^3x \,\Gamma_{\alpha\beta} \langle N_{\beta}(\vec{x},t) \overline{N}_{\alpha}(\vec{0},0)\rangle\,,
\end{equation}
with $\Gamma=(1+\gamma_0)/2$, and the ratio of the two correlation functions,
\begin{eqnarray}
R_X(t,t')=\frac{G_{{\rm 3pt},X}(t,t')}{G_{\rm 2pt}(t)}\,.
\end{eqnarray}
Performing the standard spectral decomposition of the two correlation functions and taking all times $t,t'$ and $t-t'$ to be large it is straightforward to show that the ratio $R_X$ goes to a constant.
This constant is  the forward matrix element $\langle N(\vp=0)|O_X| N(\vp=0)\rangle /2M_N$ called the nucleon charge $g_X$. In addition there are exponentially suppressed corrections from resonances and multi-hadron states that have the same quantum numbers as the nucleon. For sufficiently small pion masses the dominant multi-hadron states are two-particle nucleon-pion states with the nucleon and the pion having opposite momenta. Taking into account only these corrections 
the asymptotic behavior of the ratio reads
\begin{eqnarray}
\label{DefRatio}
R_X(t,t')= g_X\Big[1+ \sum_{\vec{p}_n} \left(b_{X,n} e^{-\Delta E_n (t-t')} + \tilde{b}_{X,n} e^{-\Delta E_n t'} + \tilde{c}_{X,n} e^{-\Delta E_n t }\right)\Big].
\end{eqnarray}
According to our assumptions about the finite spatial volume the momenta are discrete and the sum runs over all momenta allowed by the boundary conditions. $\Delta E_n = E_{N\pi,n} -M_N$ is the energy gap between the nucleon-pion state and the ground state. For weakly interacting pions $E_{N\pi,n}$ equals approximately the sum $E_{N,n}+E_{\pi,n}$ of the nucleon and pion energy.
The coefficients $b_{X,n},\tilde{b}_{X,n}$ and $\tilde{c}_{X,n}$ in \pref{DefRatio} are dimensionless ratios of various matrix elements involving the nucleon interpolating fields and the operator $O_X$.\footnote{The coefficient $\tilde{c}_{X,n}$ in the ratio  \pref{DefRatio} is proportional to the excited-to-excited-state matrix element $\langle N(\vec{p}_n)\pi(-\vec{p}_n)| O_X|N(\vec{p}_n)\pi(-\vec{p}_n)\rangle$. Contributions involving such matrix elements with different momenta in the initial and final nucleon-pion state will be ignored throughout this paper.} 
  The projection matrices $\Gamma'_X$ in table \ref{table:projectors} are chosen such that the leading constant in $R_X$ is simply the nucleon charge. Other conventions differing form ours by a factor 2 and/or a factor $i$ can be also be found in the literature. Such a choice modifies the overall constant in a trivial way, but it has no effect on the coefficients in \pref{DefRatio}.  

\subsection{The chiral effective theory}
The correlation functions defined in the previous section and their ratio can be computed in chiral perturbation theory. 
In fact, the 2-pt function has already been computed in Ref.\ \cite{Bar:2015zwa}, here we present the results for the 3-pt functions and the ratio $R_X$. We carry over the setup used for computing the 2-pt function and summarize only very briefly a few formulae. For details the reader is referred to Ref.\ \cite{Bar:2015zwa}. 

The calculations are performed to leading order in the chiral expansion in the covariant formulation of baryon ChPT \cite{Gasser:1987rb,Becher:1999he}. 
To that order the chiral effective lagrangian consists of two parts only, ${\cal L}_{\rm eff}={\cal L}_{N\pi}^{(1)} + {\cal L}_{\pi\pi}^{(2)}$.
Expanding this lagrangian in powers of pion fields and keeping interaction terms with one pion field only we obtain
\begin{eqnarray}
\label{Leff}
{\cal L}_{\rm eff} &=& \overline{\Psi} \Big(\gamma_{\mu}\partial_{\mu} +\mN \Big)\Psi +\frac{1}{2}\pi^a \Big(- \partial_{\mu}\partial_{\mu} + M_{\pi}^2 \Big)\pi^a + \frac{ig_A}{2f}\overline{\Psi}\gamma_{\mu}\gamma_5\sigma^a \Psi \, \partial_{\mu} \pi^a\,.
\end{eqnarray}
The  nucleon fields $\Psi=(p,n)^T$ and $\overline{\Psi}=(\overline{p},\overline{n})$ 
contain the Dirac fields for the proton $p$ and the neutron $n$.  $M_N,M_{\pi}$ denote the nucleon and pion masses, while $g_A$ and $f$ are the axial charge and the pion decay constant. To be precise these are the chiral limit values, but to LO they can be replaced by their values at the physical pion mass. 

The expressions for the nucleon interpolating fields in ChPT are also known \cite{Wein:2011ix}. To LO and up to one power in pion fields one finds
\begin{eqnarray}\label{Neffexp}
N(x)& = & \tilde{\alpha} \left(\Psi(x) + \frac{i}{2f} \pi^a(x)\sigma^a \gamma_5\Psi(x)\right)\,,\\
\overline{N}(0) & = & \tilde{\beta}^* \left(\overline{\Psi}(0) + \frac{i}{2f}\overline{\Psi}(0)\gamma_5\sigma^a\pi^a(0) \right)
\end{eqnarray}
These are the effective fields for local nucleon interpolating fields composed of three quarks without derivatives \cite{Ioffe:1981kw,Espriu:1983hu}. The interpolating fields do not necessarily need to be point-like, `smeared' fields map to the same chiral expressions provided two conditions are met: i) the smearing procedure is compatible with chiral symmetry and ii) the extension of the smeared fields (`smearing radius') is small compared to the Compton wavelength of the pion. In that case smeared field can be mapped onto point like fields in ChPT just like their pointlike counterparts at the quark level  \cite{Bar:2013ora,Bar:2015zwa}. Different are, however, the LECs $\tilde{\alpha},\tilde{\beta}$ entering the chiral expression in \pref{Neffexp}. If the same interpolating fields are used at both source and sink we have $\tilde{\alpha}=\tilde{\beta}$. 

For the computation of the 3-pt functions we need the expressions for the vector and axial vector currents, the scalar density and the tensor.
The first three are obtained from the known effective Lagrangian in the presence of external source fields for the currents and densities \cite{Gasser:1987rb}. Taking derivatives with respect to the external fields for vector and axial vector current we obtain from the Lagrangian ${\cal L}_{\rm eff}={\cal L}_{N\pi}^{(1)} + {\cal L}_{\pi\pi}^{(2)}$ the expressions
\begin{eqnarray}
V_{\mu}^a & = & \overline{\Psi}\gamma_{\mu}\sigma^a\Psi -\frac{g_A}{f}\epsilon^{abc} \pi^b \overline{\Psi}\gamma_{\mu}\gamma_5 \sigma^c\Psi - 2i \epsilon^{abc}\partial_{\mu}\pi^b\pi^c\,,\label{DefVector}\\
A_{\mu}^a & = & g_A  \overline{\Psi}\gamma_{\mu}\gamma_5 \sigma^a\Psi -\frac{1}{f}\epsilon^{abc} \pi^b \overline{\Psi}\gamma_{\mu}\sigma^c\Psi - 2i f\partial_{\mu} \pi^a\,.\label{DefAxial}
\end{eqnarray}
The first two terms in each expression on the right hand side stem from ${\cal L}_{N\pi}^{(1)}$, the remaining one from ${\cal L}_{\pi\pi}^{(2)}$.
For the scalar density $S^a$ we obtain  a vanishing contribution: ${\cal L}_{N\pi}^{(1)}$ does not depend on the scalar source field and the contribution from ${\cal L}_{\pi\pi}^{(2)}$ vanishes identically in SU(2) ChPT. 
The leading non-vanishing term stems from the higher order Lagrangian ${\cal L}_{N\pi}^{(2)}$. Following the notation in Ref.\ \cite{Fettes:2000gb} we obtain 
\begin{eqnarray}
S^a &=& -4 Bc_5 \overline{\psi} \sigma^a\psi\,.
\end{eqnarray}
The prefactor is a product of two LECs: $B$ is the familiar LEC proportional to the quark condensate that enters also ${\cal L}_{\pi\pi}^{(2)}$. 
The coefficient $c_5$ is a LEC in ${\cal L}_{N\pi}^{(2)}$ and has mass dimension $-1$ such that $Bc_5$ is dimensionless. To the order we are working here we will find $g_S=-4 Bc_5$, see below. 

Mesonic ChPT with a tensor source field has been constructed in Ref.\ \cite{Cata:2007ns}, but the generalization to covariant BChPT is, to our knowledge, missing. However, 
following the construction steps in Ref.\ \cite{Fettes:2000gb} it is straightforward to obtain the tensor in Baryon ChPT. Some details are summarized in appendix \ref{app:tensor}, here we just quote the final result. To leading chiral dimension we find only one term for the non-singlet tensor in ChPT, 
\begin{equation}\label{LOtensorChPT}
T^a_{\mu\nu} = - 4B c_8 \overline{\psi}\sigma_{\mu\nu}\sigma^a\psi\,.
\end{equation}
In analogy to the scalar density we have chosen to write the LEC as the product of $4B$ and an unknown LEC $c_8$ associated with the tensor field. The product is dimensionless and will be identified with the tensor charge in the next section. Obviously, the expression in \pref{LOtensorChPT} transforms as a tensor field. However, important is that this is the only tensor contributing to leading chiral dimension. 

\subsection{The 3-pt functions in ChPT}
\label{ssect3ptfunctions}
\begin{figure}[tp]
\begin{center}
\includegraphics[scale=0.5]{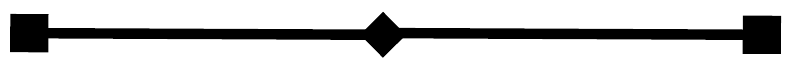}\\[3ex]
\caption{Leading Feynman diagram for the 3-pt function. Squares represent the nucleon interpolating fields at times $t$ and $0$, the diamond stands for the operator insertion at time $t'$. Solid lines represent nucleon propagators.}
\label{fig:diagNOX}
\end{center}
\end{figure}
With the expressions \pref{Leff} to \pref{LOtensorChPT} it is straightforward to compute the 3-pt functions perturbatively in ChPT. To leading order only the diagram depicted in fig.\ \ref{fig:diagNOX} leads to the single-nucleon-state contribution $G^{N}_{{\rm 3pt},X}$, and we obtain
\begin{equation}\label{SingleNucl}
G^{N}_{{\rm 3pt},X} = g_X G^{N}_{{\rm 2pt}}\,. 
\end{equation}   
$G^{N}_{{\rm 2pt}}=2\alpha\beta^* \exp{(-M_Nt)}$ denotes the leading single-nucleon-state contribution in the 2-pt function \cite{Bar:2015zwa}, and we made the identification $g_S=-4c_5B$ and $g_T=-4c_8B$ as mentioned before.

\begin{figure}[t]
\includegraphics[scale=0.45]{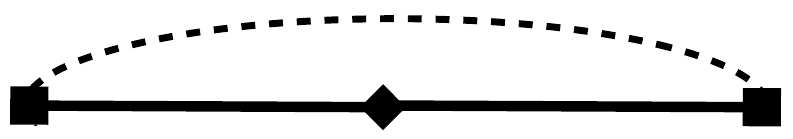}\hspace{0.3cm}\includegraphics[scale=0.45]{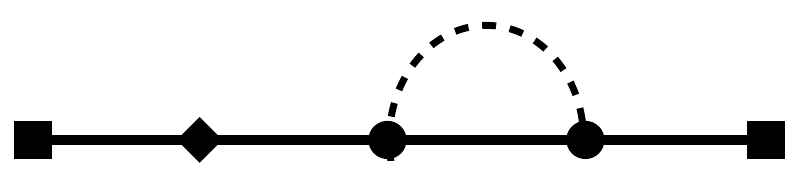}\hspace{0.3cm}\includegraphics[scale=0.45]{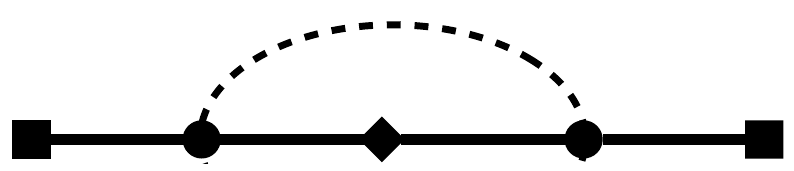}\hspace{0.3cm}\includegraphics[scale=0.45]{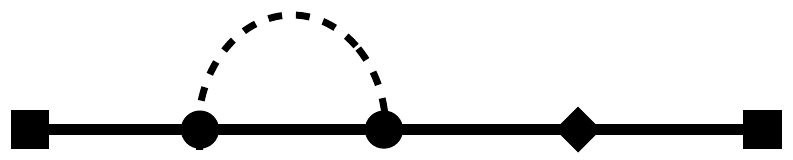}\\
a)\hspace{3.5cm} b)\hspace{3.5cm} c)\hspace{3.5cm} d)\\[3ex]
\includegraphics[scale=0.45]{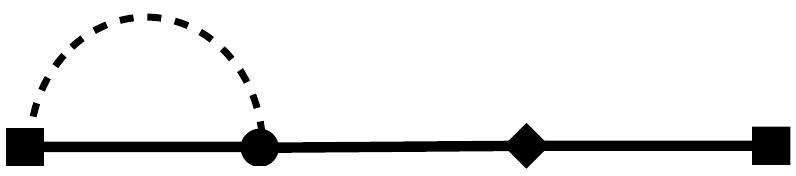}\hspace{0.3cm}\includegraphics[scale=0.45]{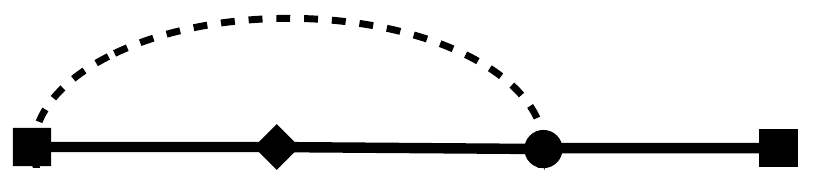}\hspace{0.3cm}\includegraphics[scale=0.45]{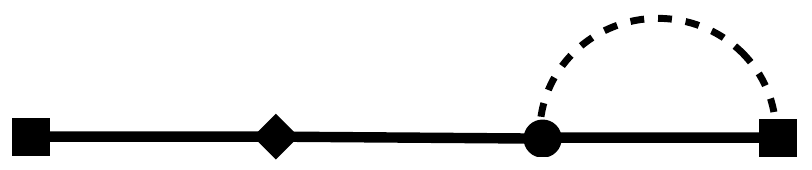}\hspace{0.3cm}\includegraphics[scale=0.45]{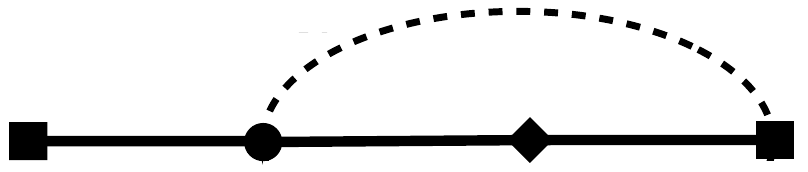}\\
e)\hspace{3.5cm} f)\hspace{3.5cm} g)\hspace{3.5cm} h) \\[3ex]
\includegraphics[scale=0.45]{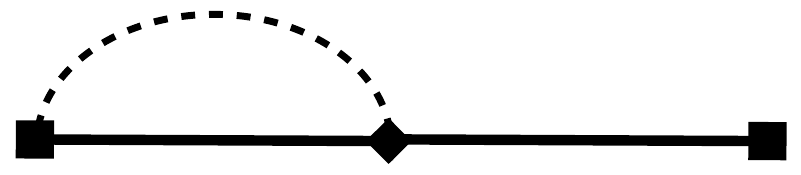}\hspace{0.3cm}\includegraphics[scale=0.45]{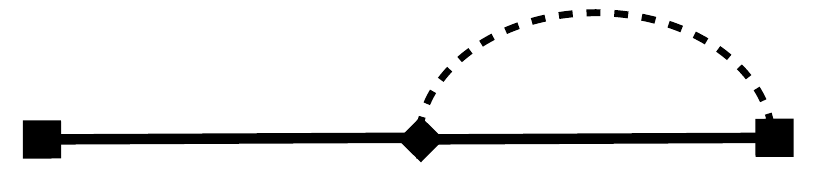}\hspace{0.3cm}\includegraphics[scale=0.45]{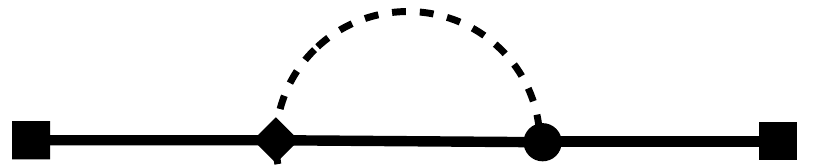}\hspace{0.3cm}\includegraphics[scale=0.45]{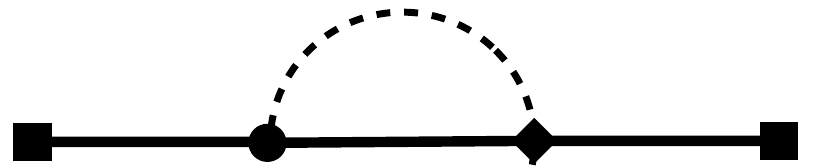}\\
i)\hspace{3.5cm} j)\hspace{3.5cm} k)\hspace{3.5cm} l)\\[3ex]
\includegraphics[scale=0.45]{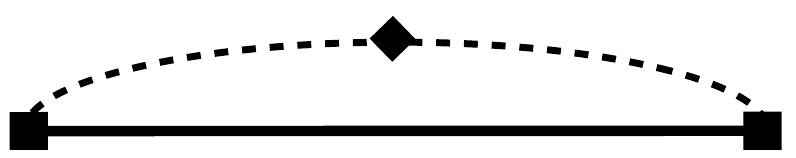}\hspace{0.3cm}\includegraphics[scale=0.45]{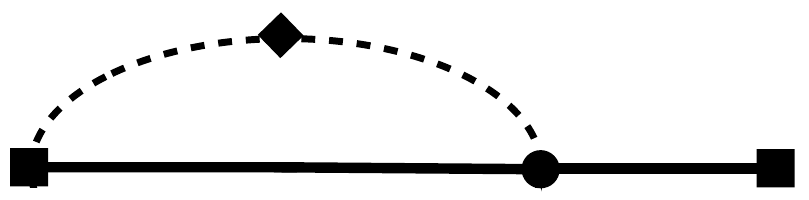}\hspace{0.3cm}\includegraphics[scale=0.45]{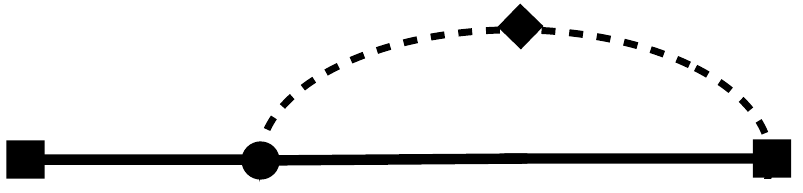}\hspace{0.3cm}\includegraphics[scale=0.45]{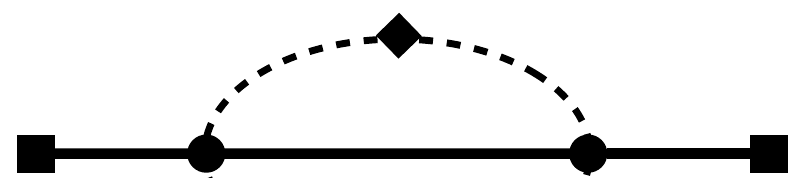}\\
m)\hspace{3.5cm} n)\hspace{3.5cm} o)\hspace{3.5cm} p)\\[0.3ex]
\caption{Feynman diagrams for the LO nucleon-pion contribution in the 3-pt functions. Circles represent a vertex insertion at an intermediate space time point, and an integration over this point is implicitly assumed. The dashed lines represent pion propagators. }
\label{fig:VCNpidiags}
\end{figure}

Figure \ref{fig:VCNpidiags} shows the diagrams with a nonzero nucleon-pion-state contribution to the 3-pt functions. Diagrams a) - h) contribute to all four correlators ($X=V,A,T,S$). In addition, diagrams i) - l) contribute to both the vector and axial vector current, while the remaining four diagrams m) - p) contribute to the vector current only. 
It will be convenient to write the nucleon-pion-state contribution $G^{N\pi}_{{\rm 3pt},X}$ in the form (we drop the subscript $n$ on the coefficients in this section)
\begin{equation}\label{NuclPionContrGeneral}
G^{N\pi}_{{\rm 3pt},X}= G^{N}_{{\rm 3pt},X} \sum_{\vec{p}_n}\left(b_{X} e^{-\Delta E_n (t-t')} + \tilde{b}_{X} e^{-\Delta E_n t'}  + c_{X} e^{-\Delta E_n t} \right).
\end{equation}
As already mentioned, vector current conservation implies that the 3-pt function is given by the conserved charge times the 2-pt function. In terms of the coefficients in \pref{NuclPionContrGeneral} this statement reads
\begin{equation}\label{WTIV0}
b_{V_0}=\tilde{b}_{V_0}=0\,,\quad c_{V_0}=c_{{\rm 2pt}}\,.
\end{equation}
We have checked this result explicitly, and it provided a non-trivial test on the programs we have written to compute the diagrams for general fields $O_X$ and $\Gamma'_X$. 

To quote the results for $X=A,T,S$ we introduce some short hand notation. Since some overall factors are common to all coefficients we write the coefficients according to 
\begin{eqnarray}
b_{X} &=& \frac{1}{16(fL)^2E_{\pi}L} \left(1-\frac{M_N}{E_N}\right)B_{X}\,,\label{finalb}\\
c_{X} &=& \frac{1}{16(fL)^2E_{\pi}L} \left(1-\frac{M_N}{E_N}\right)C_{X}\,.\label{finalc}
\end{eqnarray}
In our calculation we explicitly found 
\begin{equation}\label{eqcoeff}
\tilde{b}_{X}=b_{X}
\end{equation}
for all correlators, so we need to quote only $b_{X}$. Note that the coefficients vanish if the momentum of the nucleon (and the pion) is zero. This has to be the case since the nucleon-pion state with both particles at rest does not contribute to the correlators for symmetry reasons. 

For the `reduced' coefficients $C_X$ we find the following results:
\begin{eqnarray}
C_{A} &=& \left(\gAb -1\right)^2 \frac{2}{3}\left(\frac{M_N}{E_N}-\frac{1}{2}\right)\,, \\ 
C_{T} &=& \left(\gAb -1\right)^2 \frac{1}{3}\left(2-\frac{M_N}{E_N}\right)\,,\\
C_{S} &=& \left(\gAb -1\right)^2 \left(-\frac{M_N}{E_N}\right)\,.
\end{eqnarray}
For notational simplicity only we have introduced the combination 
\begin{equation}
\gAb=g_A\frac{E_{N\pi} + M_N}{E_{N\pi} -M_N}\,, \quad E_{N\pi}=E_N+E_{\pi}\,,
\end{equation}
which appears also in the results for the coefficients $B_X$:
\begin{eqnarray}
B_{A} &=& \frac{8}{3}\left(\gAb -1\right) \left( \gAb -\frac{1}{2} \gA \frac{M_{\pi}^2}{2E_{\pi}M_N-M_{\pi}^2}\right) - 4\left(\frac{E_{N\pi}+M_N}{E_{N\pi}-M_N} -\frac{1}{g_A}\right)\,,\\
B_{T} &=& \frac{8}{3}\left(\gAb -1\right) \left( \gAb +\frac{1}{4} \gA \frac{M_{\pi}^2}{2E_{\pi}M_N-M_{\pi}^2}\right)\,,\\
B_{S} &=& 4\left(\gAb -1\right) \left( \gAb +\frac{1}{2} \gA \frac{M_{\pi}^2}{2E_{\pi}M_N-M_{\pi}^2}\right) \,.
\end{eqnarray}
The axial vector correlation function was also calculated in Ref.\ \cite{Tiburzi:2015tta} using heavy baryon (HB) ChPT. If we expand $E_N\sim M_N+p^2/2M_N$ in our result for the axial vector current and drop all but the dominant terms we do reproduce the result in Ref.\ \cite{Tiburzi:2015tta}.

Taking the ratio of the 3-pt and 2-pt function we find $R_X$ given by the form anticipated in \pref{DefRatio}, with the coefficients 
\begin{eqnarray}
\tilde{c}_{X} = c_X - c_{\rm 2pt}\,.
\end{eqnarray}
The coefficient stemming from the 2-pt function reads \cite{Bar:2015zwa}
\begin{eqnarray}
c_{\rm 2pt} &=& \frac{1}{16(fL)^2E_{\pi}L} \left(1-\frac{M_N}{E_N}\right)C_{\rm 2pt}\,,\quad C_{\rm 2pt} \,=\, 3\left(\gAb -1\right)^2\,.
\end{eqnarray}
The coefficients $b_X,\tilde{c}_X$ depend on two LECs only, $f$ and $g_A$, the coefficients $B_X,\tilde{C}_X$ depend only on $g_A$. 
The LECs associated with the interpolating field, on the other hand, cancel in the ratio. Thus, the LO result we have found here is universal and applies to pointlike and smeared interpolating fields. However, at the next order in the chiral expansion this universality property will be lost.
 
The ratios $b_X/b_{X'}$ and $\tilde{c}_X/\tilde{c}_{X'}$ are related and depend only on $g_A$. Since $g_A$ is known rather well from phenomenology our LO calculation makes concrete predictions for the relative size of the nucleon-pion-state contributions. These relations are particularly simple in the HB limit, where we find the equality 
\begin{equation}
b^{HB}_{A} = - \tilde{c}^{HB}_{A}
\end{equation}
for the coefficients in the axial vector case and,  in addition, 
\begin{eqnarray}\label{HBrelations}
b^{HB}_{A} &=& b^{HB}_{T} \,=\, \frac{2}{3} b^{HB}_{S}\,,\quad \tilde{c}^{HB}_{A} \,=\, \tilde{c}^{HB}_{T} \,=\, \frac{2}{3} \tilde{c}^{HB}_{S}\,,
\end{eqnarray}
relating them to the coefficients for the tensor and scalar. We would thus conclude that the nucleon-pion-state contributions are equal for the axial vector and the tensor, and fifty percent larger for the scalar. Away from the heavy baryon limit the simple relations \pref{HBrelations} will be modified, see next section.

A final comment concerns the summation over the lattice momenta in \pref{DefRatio}. Momenta that are related by the symmetries of the spatial lattice lead to the same contribution, hence it is convenient to  sum over the absolute value $p_n=|\vec{p}_n|$. Imposing periodic boundary conditions the absolute value can assume the values $p_n=(2\pi/L)\sqrt{n}$, $n\equiv n_1^2+n_2^2+n_3^2$, with the $n_k$ being integers. Therefore, in the ratio we can perform the replacement 
\begin{equation}\label{redsum}
\sum_{\vec{p}} \longrightarrow \sum_{p_n} m_n\,,
\end{equation}
where the multiplicities $m_n$ count the number of vectors $\vec{p}_n$ with the same $p_n$. Multiplicities for $n\leq 20$ are given in Ref.\ \cite{Colangelo:2003hf} (for convenience we summarize the first eight in table \ref{tabledn}).

\begin{table}[tbp]
\begin{center}
\begin{tabular}{lrrrrrrrrr}
\hline\hline
$n$ & 0 & 1 & 2 & 3 & 4 & 5 & 6 & 7 & 8\\ 
$m_n$ & \phantom{1}1 & \phantom{1}6 & 12 & \phantom{1}8 & \phantom{1}6 & 24 & 24 & \phantom{0}0 & 12\\
\hline
\end{tabular}
\end{center}
\caption{\label{tabledn}Multiplicities $m_n$ in eq.\ \pref{redsum} for $n\le 8$ (see Ref.\ \cite{Colangelo:2003hf}).}
\end{table}

\section{Impact on lattice calculations of the nucleon charges}

\subsection{Preliminaries}

In the following we want to estimate the impact of the nucleon-pion-state contribution on the determination of the various charges in lattice QCD simulations. Two methods are widely used, the plateau and the summation method. Before considering them in the next two sections a few preliminary remarks need to be made.

Our result for the ratio $R_X$ can be written as
\begin{eqnarray}\label{simplnotationR}
R_X(t,t') &=& g_X\Big[1+ \sum_{n\le n_{\rm max}} b_{X,n}\left(e^{-\Delta E_n(t-t')} + e^{-\Delta E_n t'}\right) + \tilde{c}_{X,n} e^{-\Delta E_n t} \Big]\,,
\end{eqnarray}
where we used eq.\ \pref{eqcoeff}. The coefficients $b_{X,n},\tilde{c}_{X,n}$ are dimensionless and depend on four independent dimensionless parameters: $\gA, f/M_N, M_{\pi}/M_N$ and $M_{\pi}L$. To leading order in the chiral expansion we can use the physical values for the two LO LECs, i.e.\ we set  $\gA=1.27$ and $f=f_{\pi}= 93$ MeV. Since we are mainly interested in $R_X$ for physical pion masses we fix  the pion and nucleon mass to their physical values, thus we take $M_{\pi}/M_N=140/940$ and $f/M_N=93/940$ if not stated otherwise. 

The ratio $R_X$ also depends on $n_{\rm max}$, the upper limit for the number of states taken into account in the ratio. 
In ChPT $n_{\rm max}$ is essentially determined by insisting on a sufficiently small expansion parameter $p_n/\Lambda_{\chi}$ in (finite volume) ChPT, with $\Lambda_{\chi}$ typically identified with $4\pi f_{\pi}$ \cite{Colangelo:2003hf}.  In Ref.\ \cite{Bar:2015zwa}  the condition $p_{n_{\rm max}}/\Lambda_{\chi}= 0.3$ was imposed for a reasonably well behaved chiral expansion, and we adopt this choice in the following as well. This bound translates into $n_{\rm max} =2$ and 5 for $M_{\pi}L=4$ and 6, respectively. A second reason for this particular bound is that the energy $E_{N\pi,{n_{\rm max}}}$ of the nucleon-pion-states satisfying it is sufficiently well below the energy of the first resonance state with an expected energy of about 1.5$M_N$. In that case we may ignore mixing effects with this resonance state that is not included as a degree of freedom in the chiral effective theory. 

\begin{table}[tp]
\begin{center}
\begin{tabular}{l|c|c|c}
\multirow{2}{*}{$\frac{p_{n_{\rm max}}}{\Lambda_{\chi}}$}& \multicolumn{2}{|c|}{$n_{\rm max}$ } &  \multirow{2}{*}{$\frac{E_{N\pi,{n_{\rm max}}}}{M_N}$} \\ 
& $M_{\pi}L=4$ & $M_{\pi}L=6$ & \\  \hline
0.3 & 2 & 5 & $\approx 1.35 $ \\
0.45 & 5 & 12 & $\approx 1.6\phantom{3}$\\
0.6 & 10 & 22 & $\approx 1.9\phantom{3}$
\end{tabular}
\end{center}
\caption{$n_{\rm max}$ and $E_{N\pi,{n_{\rm max}}}$ as a function of $p_{n_{\rm max}}/\Lambda_{\chi}$, see main text.}
\label{table:nmax}
\end{table}
Obviously there is some arbitrariness in imposing a bound on the momenta and the values for $n_{\rm max}$ following from it.
In the end $n_{\rm max}$ must be large enough such that the contribution from the states omitted in the ratio $R_X$ is small enough that it can be ignored. This depends essentially on the times $t$ and $t'$ that govern the exponential suppression in $R_X$. In  table \ref{table:nmax} we have collected three examples for bounds on the momentum and the associated values $n_{\rm max}$. Two of the bounds imply energies $E_{N\pi,{n_{\rm max}}}$ above the energy of the first resonance state. Going to such high energies will give some indication about the impact of the nucleon-pion-states, still, as long as the resonance is not included in the effective theory the results should be interpreted with care.

\subsection{Impact on the plateau method}

The excited-state contribution in $R_X$ is minimal for the operator insertion time in the middle between source and sink. Thus we may take the 'midpoint' value $R_X(t,t/2)$ as an estimate for the nucleon charge $g_X$. This midpoint method is essentially equivalent to what is called 'plateau method', so we will use this terminology here as well. 

\begin{figure}[p]
\begin{center}
\mbox{$R_X/g_X$}\\
\includegraphics[scale=0.9]{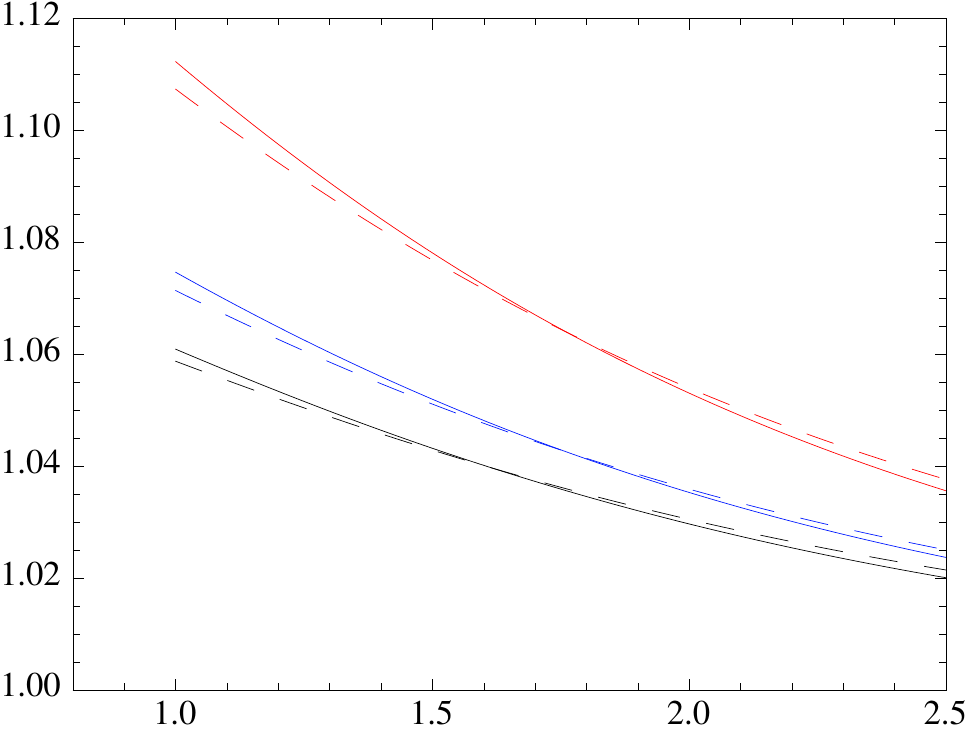}\\
\mbox{$t$ [fm]}
\caption{The plateau estimate $R_X(t,t/2)$ normalized by $g_X$ for all three charges ($X=A$ in black, $T$ in blue, $S$ in red). Results for $M_{\pi}=140$ MeV and for $M_{\pi}L=4$ (solid lines) and $M_{\pi}L=6$ (dashed lines). $n_{\rm max}$ according to the first row in table \ref{table:nmax}.}
\label{fig:midpoint140}
\end{center}
\end{figure}

Figure \ref{fig:midpoint140} shows $R_X(t,t/2)/g_X$, the plateau method estimate divided by the charge. Without the $N\pi$ contribution this ratio would be equal to 1, and the deviation from this value is the relative error in percent caused by the $N\pi$ contribution. Plotted are the results for all three charges ($X=A,T,S$) for two values of $M_{\pi}L$ (4 and 6). The following observations can be made:\\
(i) The differences between the results for $M_{\pi}L=4$ and $M_{\pi}L=6$ are very small. These differences stem from the fact that the energy interval of the nucleon pion states that we consider, $[M_N+M_{\pi}, E_{{N\pi},n_{\rm max}}]$, contains only 2 and 5 states for $M_{\pi}L=4$ and 6, respectively. In infinite volume there will be states to any energy, so some finite volume effect in the nucleon-pion-state contribution is expected. Still, it is perhaps somewhat surprising that the differences between $M_{\pi}L=4$ and $M_{\pi}L=6$ are so small.\\
(ii) The results for the axial vector and the tensor charge are very close, and the result for the scalar charge is about 50\% larger. This is in good agreement with the expectation \pref{HBrelations} for the coefficients in the heavy baryon limit.\\ 
(iii) All three curves in figure are above 1, so the nucleon-pion-state contribution leads to an overestimation of the three charges.

%
\begin{figure}[p]
\begin{center}
\mbox{$R_X/g_X$}\\
\includegraphics[scale=0.90]{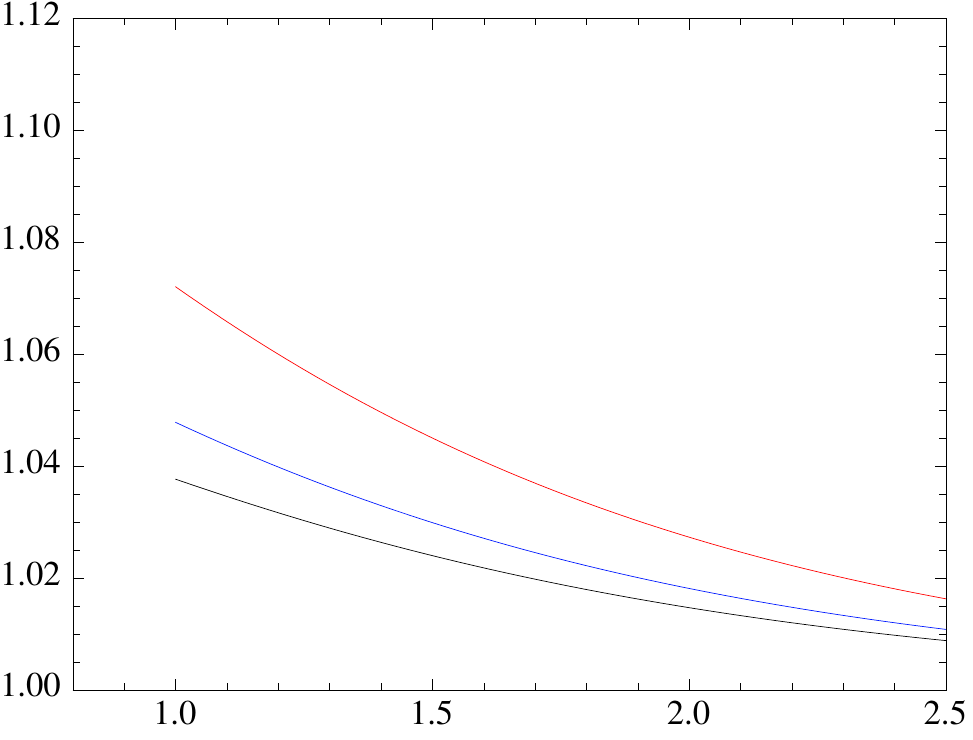}\\
\mbox{$t$ [fm]}
\caption{The plateau estimate $R_X(t,t/2)$ normalized by $g_X$ for all three charges ($X=A,T,S$, same color code as in figure \ref{fig:midpoint140}).  Results for $M_{\pi}=200$ MeV, $M_{\pi}L=4$ and $n_{\rm max}=1$.}
\label{fig:midpoint200v2}
\end{center}
\end{figure}

\begin{figure}[p]
\begin{center}
\mbox{$R_A/g_A$}\\
\includegraphics[scale=0.9]{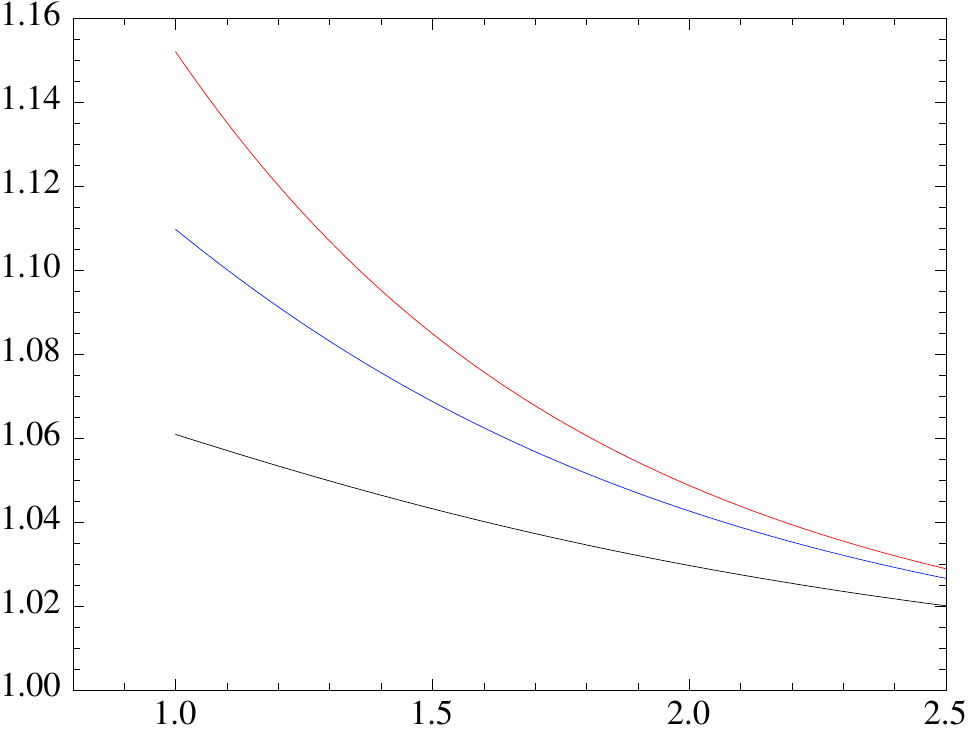}\\
\mbox{$t$ [fm]}
\caption{The plateau estimates $R_A(t,t/2)$ normalized by $g_A$ for $M_{\pi}=140$ MeV, $M_{\pi}L=4$ and the three different $n_{\rm max}$ values specified in table \ref{table:nmax} ($n_{\rm max}=2$ in black, 5 in blue and 10 in red).}
\label{fig:midpointA3nmax}
\end{center}
\end{figure}

\begin{figure}[p]
\mbox{$R_S/g_S$}\\
\includegraphics[scale=0.9]{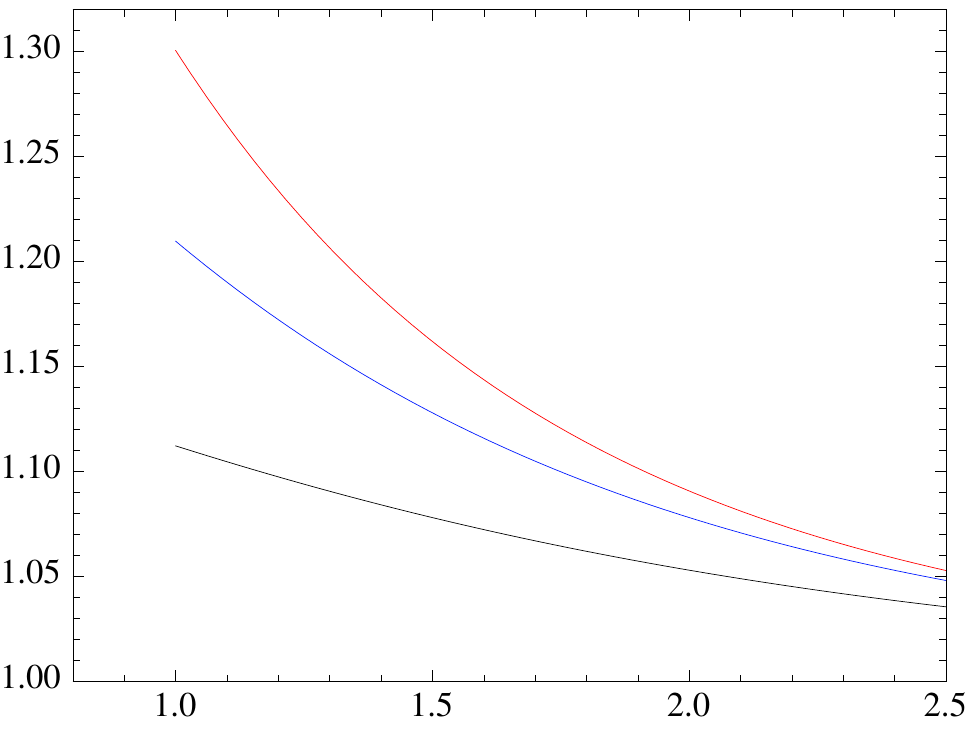}\\
\mbox{$t$ [fm]}
\caption{The plateau estimates $R_S(t,t/2)$ normalized by $g_S$ for $M_{\pi}=140$ MeV, $M_{\pi}L=4$ and the three different $n_{\rm max}$ values specified in table \ref{table:nmax} (same color code as in fig.\ \ref{fig:midpointA3nmax}).}
\label{fig:midpointS3nmax}
\end{figure}
As an illustration of the pion mass dependence figure \ref{fig:midpoint200v2} shows the results for the pion mass $M_{\pi}=200$ MeV and $M_{\pi}L=4$. In this case the bound $p_{n_{\rm max}}/\Lambda_{\chi}= 0.3$ on the momenta leads to $n_{\rm max}=1$, so only the nucleon-pion state with the smallest non-zero momentum is taken into account. While the nucleon-pion-state contribution for the axial and the tensor charge  are still roughly the same and smaller compared to the scalar, the absolute size is about a factor one half smaller compared with the results for the physical pion mass.

Figure \ref{fig:midpointA3nmax} shows the dependency of the results on $n_{\rm max}$
for the axial vector ratio. The results are shown for the three $n_{\rm max}$ values specified in table \ref{table:nmax} and for $M_{\pi}L=4$ (the counterparts for $M_{\pi}L=6$ lie essentially on top of the curves in figure \ref{fig:midpointA3nmax}). 
The result for the lowest $n_{\rm max}$ starts to be the dominant part of the $N\pi$ contribution at about 2 fm. Recall that the smallest $n_{\rm max}$ corresponds to the lower tail of the nucleon-pion states that ends below the first resonance. Apparently, this lower tail does not capture properly the $N\pi$ contribution at and below 1.5 fm where it only makes about one half or even less of the $N\pi$ contribution with the largest $n_{\rm max}$. 

Figure \ref{fig:midpointA3nmax} tells an important message: Unless the source- sink separation is larger than about 1.5 fm the nucleon-pion states with energies above the first resonance state contribute significantly to the ratio.
 Therefore, the impact of this resonance needs to be included before definite conclusions about the overall excited-state-contamination in the ratio can be drawn. Still, unless there are large cancellations caused by the resonance state we may estimate the $N\pi$ contribution to $g_A$ to be at the +5\% to +10\% level. 

Figure \ref{fig:midpointS3nmax} is the analogous plot for the ratio $R_S$. It looks qualitatively the same as figure \ref{fig:midpointA3nmax}, but the size of the corrections is about twice as large compared to the axial vector case. The $N\pi$ contribution to $R_T$ (not shown) is  about 25\% larger than the corresponding one to $R_A$.  

\subsection{Impact on the summation method}
Suggested originally in Ref.\ \cite{Maiani:1987by} the summation method was first applied in Ref.\ \cite{Capitani:2012gj} in the determination of  $\gA$. The main observation underlying this method is that the ratio $R_X(t,t')$ apparently has a stronger exponential suppression once the sum over all insertion times $t'$ is taken. The asymptotic behavior anticipated in \cite{Capitani:2012gj} reads\footnote{This is a lattice QCD formula. The summation is over the discrete operator insertion times and $g_A$ denotes the bare axial charge.} 
\begin{equation}\label{indmethodgeneral}
S_A(t)\equiv \sum_{t'=0}^t R_A(t,t') \, \longrightarrow g_A[1+ {\rm O}(e^{-\Delta E t})] t + {\rm const.} + {\rm O}(e^{-\Delta E t})\,.
\end{equation}
Here $\Delta E$ denotes the energy gap between the ground and first excited state. Without the excited-state contribution the sum shows a simple linear $t$ dependence with the slope given by the charge. The presence of excited states results in exponentially suppressed corrections. 
In practice the slope is obtained by fitting a linear function to lattice data for various sink times $t$. 

With the results for the $N\pi$ contribution to the ratio $R_X(t,t')$ we can study their impact on the summation method. Since our underlying space time manifold here is continuous the sum in \pref{indmethodgeneral} is replaced by the integral and the slope can be computed directly by taking the time derivative. However, one caveat needs to be kept in mind: $S_X(t)$ involves the 3-pt function at short time differences $t-t'$ and $t'$, and these are not properly captured by the chiral effective theory. Even though we can compute the nucleon-pion-state contribution to $S_X(t)$ it is unclear how much their contribution is distorted by the short distance contributions to $S_X(t)$. 

That being said,  we consider the generalized sum $S_X(t,\tm)$ introduced in Ref.\ \cite{Bali:2014nma}, where the sum (integral) over $t'$ is taken over the interval $[\tm,t-\tm]$, with $\tm \le t/2$. For $\tm$ sufficiently large the nucleon-pion-state contribution is expected to give the dominant excited-state correction to $S_X(t,t_{\rm m})$, and it can be computed within ChPT.  In the end we can send $\tm$ to zero bearing in mind the caveat mentioned before.

With the result for the ratio $R_X$ in \pref{simplnotationR} the integral $S_X(t,t_{\rm m}) = \int^{t-\tm}_{\tm} dt' R_X(t,t')$ reads
\begin{equation}
S_X(t,\tm) =g_X\left[ \left(1+ \sum_{n\le n_{\rm max}} \tilde{c}_n e^{-\Delta E_n t}\right) (t-2\tm) +  \sum_{n \le n_{\rm max}}  \frac{2b_n}{\Delta E_n} \left(e^{-\Delta E_n \tm} - e^{-\Delta E_n (t-\tm)}\right)\right].\label{integratedRatio}
\end{equation}
Setting $\tm$ equal to zero we do recover the $t$ dependence in \pref{indmethodgeneral}. As a function of $t$ (keeping $\tm$ fixed) the slope $s_X(t,\tm)\equiv dS_X(t,\tm)/dt$ is given by
\begin{eqnarray}\label{slope}
s_X(t,\tm) &=& g_X\left[ 1+ \sum_{n\le n_{\rm max}} \!\tilde{c}_n\{1-\Delta E_n (t-2\tm)\}  e^{-\Delta E_n t} + \sum_{n\le n_{\rm max}} 2b_n e^{-\Delta E_n (t-\tm)} \right].
\end{eqnarray} 
Note that the dependence of $s_X(t,\tm)$ on $\tm$  decreases the larger $t$ is, and it vanishes in the infinite $t$ limit, as expected.
 
\begin{figure}[p]
\mbox{$s_A/g_A$}\\
\includegraphics[scale=0.9]{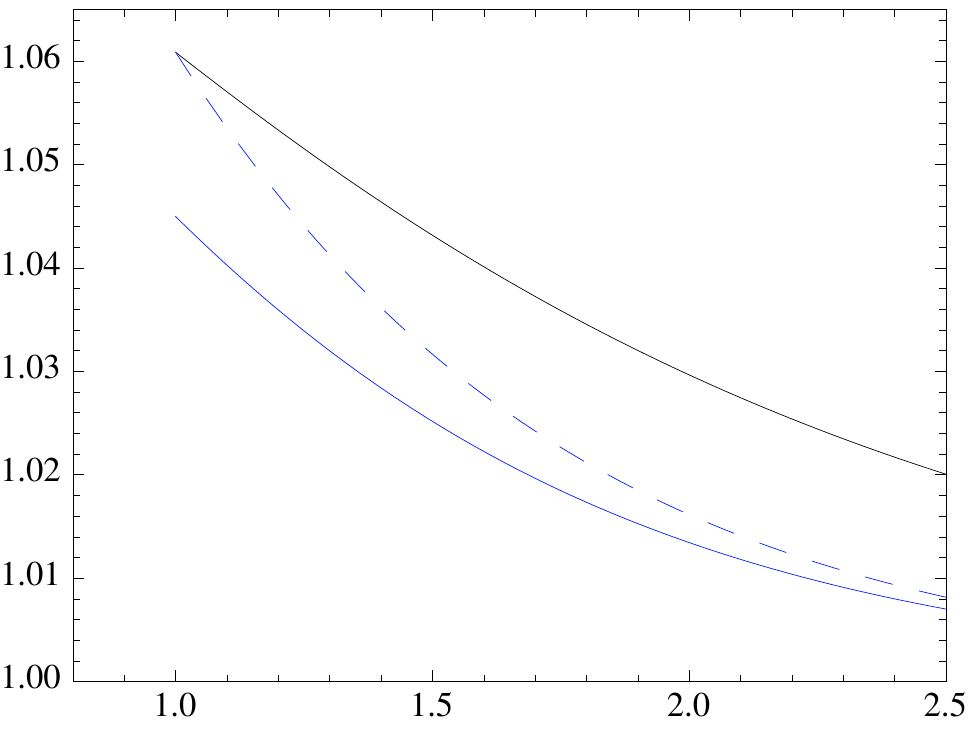}\\
\mbox{$t$ [fm]}
\caption{The summation method estimate $s_A(t,t_m)$ normalized by $g_A$ for $M_{\pi}=140$ MeV, $M_{\pi}L=4$ and $t_m=0$ (solid blue line) and $t_m=0.5$ fm (dashed blue line). For comparison the plateau method estimate $R_A(t,t/2)/g_A$ is also shown (black solid line).
}
\label{fig:compsumplat}
\end{figure}

Figure \ref{fig:compsumplat} shows $s_A(t,\tm)/g_A$ for $\tm=0.5$ fm and $t>2\tm$. We have chosen this value to admit a comparison with the plateau method result, which is also plotted in figure \ref{fig:compsumplat}. Note that for $\tm=t/2$ both methods agree since $s_A(t,t/2)=R_A(t,t/2)$. For $t>\tm$, however, the $N\pi$ contribution decreases more rapidly for the summation method due to the suppression caused by the exponentials $\exp[-\Delta E_n(t-\tm)]$. 

\begin{figure}[p]
\mbox{$s_A/g_A$}\\
\includegraphics[scale=0.9]{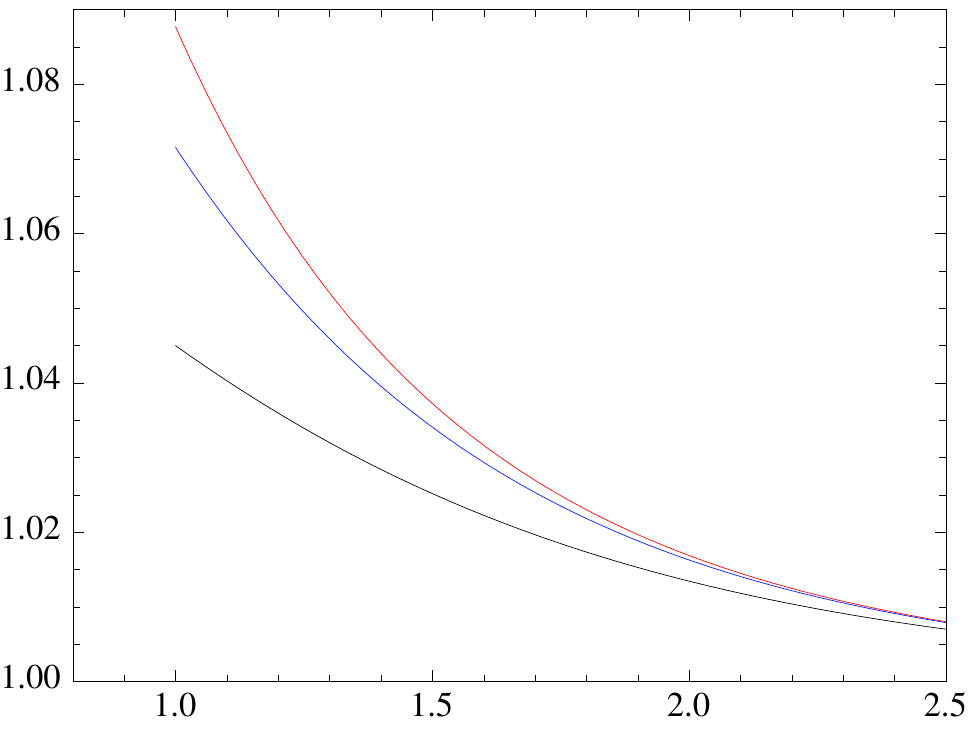}\\
\mbox{$t$ [fm]}
\caption{The summation method estimate $s_A(t,0)$ normalized by $g_A$ for $M_{\pi}=140$ MeV, $M_{\pi}L=4$ and the three different $n_{\rm max}$ values specified in table \ref{table:nmax} ($n_{\rm max}=2$ in black, 5 in blue and 10 in red).}
\label{fig:sum3nmax}
\end{figure}

The results look qualitatively the same if $\tm$ is changed. The result for $\tm\rightarrow 0$ is also shown in figure \ref{fig:compsumplat},  and the curves corresponding to $\tm$ between 0 and 0.5 fm lie between the two curves shown in the figure.
Two main conclusions  can be drawn from these results: \\
(i) The $N\pi$ contribution leads to an overestimation of the axial charge since $s_A(t,\tm)/g_A$ is larger than one. The larger $\tm$ the larger the overestimation, even though the dependence on $\tm$ vanishes rapidly.\\
(ii) The $N\pi$ contribution to the summation method is smaller compared to the plateau method. How much smaller depends on $t$, but for the range covered in the figure the summation method estimate is about 30\% to 60\% smaller than the plateau estimate.

Figure \ref{fig:sum3nmax} shows the dependence on $n_{\rm max}$. Not surprisingly, we find the same qualitative behavior as for the plateau method, cf.\ figure \ref{fig:midpointA3nmax}. However, the lower tail of the $N\pi$ contribution ($n_{\rm max}=2$) forms the dominant part of the entire contribution at significantly smaller sink times. 

The same observations can be made for the tensor and scalar charges. The results are qualitatively the same as in figures \ref{fig:compsumplat} and \ref{fig:sum3nmax}, but the size of the $N\pi$ correction is slightly larger for the tensor and about 50\% larger for the scalar. 

One needs to be careful in drawing conclusions from the results found here to actual lattice QCD data. As mentioned before, in practice the derivative with respect to sink time is obtained by a linear fit to data for sink times with finite differences. In addition, the statistical errors are usally much smaller for the data at small sink times. Thus, the fit can be significantly weighted by the data for the smallest source-sink separation \cite{Dragos:2016rtx} and may match the slope at the smallest sink time used in the fit.

Moreover, knowing the nucleon-pion-state contribution to the slope for vanishing $\tm$ might be of limited use since the short distance contributions to the 3-pt function may have a significant impact on the slope. An observation in support of this is the following: Eq.\ \pref{integratedRatio} seems to suggest that the dominant $N\pi$ contribution to the slope stems from the $\tilde{c}_{X,n}$ correction, since this contribution modifies directly the prefactor of $(t-2\tm)$. This, however, is not the case. 
The $b_{X,n}$ contribution in \pref{slope} dominates the slope, but this is also the contribution that will be affected by the short distance contribution not included in ChPT.

\section{Concluding remarks}

Some collaborations have already performed lattice simulations of the various nucleon charges on ensembles with a pion mass at or near the physical value \cite{Bali:2014nma,Abdel-Rehim:2015owa,Bhattacharya:2016zcn,vonHippel:2016wid}. Applying the conclusions found here to these numerical results is hampered mainly by the small source-sink separations $t$ in these simulations. In most cases the maximal source-sink separation $t_{\rm max}$ is about 1.2 fm, sometimes even smaller, but in all cases not much above 1.5 fm. 

As we have seen, for such small source-sink separations nucleon-pion states with energies up to about twice the nucleon mass contribute  significantly to the ratios $R_{X}$. This uncomfortably high value is way above the energy of the first resonance states.  These were not included as degrees of freedom in our chiral effective theory, but presumably these states have a non-negligible contribution to the ratios at small $t$. 
Some qualitative features of our results may still survive the omission of the resonances (overestimation of all charges by both the plateau and the summation method, a larger $N\pi$ contribution in the scalar charge), but this is not guaranteed.

On the other hand, the calculation presented here can be improved to remedy its limitations. A way to include the Roper resonance in the chiral effective theory has been known for some time \cite{Borasoy:2006fk}. The $\Delta$ resonance too can be incorporated in the effective theory \cite{Djukanovic:2009gt,Bauer:2012at,Gegelia:2016xcw}. With these additional dynamical degrees of freedom in the theory one may expect to be able to assess the excited-state contributions to the nucleon charges at much smaller source-sink separations with smaller and controllable errors. Whether contact with present day lattice simulations can be made remains to be seen though. Obviously, lattice simulations with larger source-sink separations than used today would help in this respect.

Compilations of the numerous lattice calculations of $g_A$ for larger than physical pion masses can be found in various recent reviews \cite{Green:2014vxa,Syritsyn:2014saa,Constantinou:2015agp}. 
In almost all cases the lattice estimate is smaller than the experimental value. This underestimation is more pronounced for heavier pion masses and seems to ease for $M_{\pi}$ approaching its physical value.  Whether the $N\pi$ contribution plays some role in this cannot be said for sure. Still, the possibility that a diminishing discrepancy with the experimental value is caused by more than one source of error that partially cancel each other for a decreasing pion mass should not be discarded right away.

\vspace{2ex}
\noindent {\bf Acknowledgments}
\vspace{2ex}

I thank Jeremy Green for discussions on the Ward identity for the vector current and Akaki Rusetsky for pointing out references on the tensor in ChPT. I also thank the Yukawa Institute for Theoretical Physics for its kind hospitality. This work is supported by the Japan Society for the Promotion of Science (JSPS) with an Invitation Fellowship for Research in Japan (ID No. L16520).
\vspace{3ex}

\begin{appendix}

\section{The tensor field in Baryon ChPT}\label{app:tensor}
Mesonic chiral perturbation theory with a tensor source field has been constructed in Ref.\ \cite{Cata:2007ns}. Generalizing the familiar procedure employed by Gasser and Leutwyler in Ref.\ \cite{Gasser:1983yg} a source term for the tensor field is added to the massless QCD lagrangian. This source term is mapped to ChPT taking into account its transformation properties under chiral symmetry, parity and charge conjugation. 

In terms of chiral fields the source term has the form\footnote{In this appendix we assume the Minkowski space-time metric in order to match the conventions in Refs.\  \cite{Cata:2007ns,Fettes:2000gb}.} 
\begin{equation}\label{Lexttensor}
{\cal L}_{\rm tensor}=\overline{\psi}_R t_{\mu\nu}\sigma^{\mu\nu}\psi_L +  \overline{\psi}_L t^{\dagger}_{\mu\nu}\sigma^{\mu\nu}\psi_R 
\end{equation}
with the matrix valued source field $t_{\mu\nu}$. It couples left- and right handed fields like the source term $\chi$ involving the scalar and pseudoscalar densities. Under chiral transformations $R,L$ the source term is invariant if the source field transforms according to  
$t_{\mu\nu} \longrightarrow Rt_{\mu\nu}L^{\dagger}$, $t^{\dagger}_{\mu\nu} \longrightarrow L t^{\dagger}_{\mu\nu} R^{\dagger}$. Similarly, the tensor source field needs to be even under parity and odd under charge conjugation for \pref{Lexttensor} to be invariant under these transformations as well. Based on these symmetry properties the source term can be mapped to ChPT. Postulating the power counting $t_{\mu\nu} \sim {\rm O}(p^2)$ the leading terms start at ${\rm O}(p^4)$ since at least two derivatives are needed to form a Lorentz scalar with the tensor source. The complete lagrangian through  ${\rm O}(p^6)$ can be found in \cite{Cata:2007ns}.

For the construction of the chiral lagrangian in Baryon ChPT following Ref.\ \cite{Fettes:2000gb} it is useful to introduce the combinations
\begin{equation}\label{Deftmunupm}
t_{\mu\nu,\pm} = u^{\dagger} t_{\mu\nu}u^{\dagger} \pm u t^{\dagger}_{\mu\nu}u\,.
\end{equation}
with $u$ being the standard chiral field containing the pion fields. The reason for this definition is that these fields transform as all the other external source fields under chiral symmetry, namely $t_{\mu\nu,\pm} \longrightarrow h t_{\mu\nu,\pm} h^{-1}$, where $h$ denotes the compensator field associated with the non-linear realization of chiral symmetry \cite{Coleman:1969sm,Callan:1969sn}. 

Invariants under chiral symmetry are therefore easily constructed. Following section 2.2.\ of Ref.\ \cite{Fettes:2000gb} any invariant monomial in the effective $N\pi$ Lagrangian is of the generic form
\begin{equation}
\overline{\psi}A^{\mu\nu\ldots} \Theta_{\mu\nu\ldots} \psi + {\rm h.c.}\,\,.
\end{equation}
Here $A^{\mu\nu\ldots}$ is a product of pion and/or external fields and their covariant derivatives, while $\Theta_{\mu\nu\ldots}$ is a product of a Clifford algebra element and a totally symmetrized product of covariant derivatives acting on the nucleon fields. These objects obey various restrictions stemming from chiral symmetry. In addition, equations of motion can be used to remove terms in the chiral lagrangian that are redundant. 

Here we are interested only in the leading terms involving the tensor source field only once. The simplest terms with lowest chiral dimension are obtained with $A^{\mu\nu} = t^{\mu\nu}_+$. Since the tensor source is antisymmetric in the Lorentz indices there is only one independent term $\Theta_{\mu\nu} = \sigma_{\mu\nu}$ one can contract $A^{\mu\nu}$ with. Therefore, to leading chiral dimension the external source term \pref{Lexttensor} is mapped onto 
\begin{equation}\label{LexttensorChPT}
{\cal L}^{(2)}_{\rm tensor} = c_8 \overline{\psi} t^{\mu\nu}_+ \sigma_{\mu\nu} \psi + c_9 \overline{\psi} \langle t^{\mu\nu}_+\rangle  \sigma_{\mu\nu} \psi\,.
\end{equation}
Taking the derivative with respect to the tensor source field and expanding in powers of pion fields it is straightforward to derive the expression \pref{LOtensorChPT} for the tensor field. 

The power counting for the tensor source term deserves a comment. We assumed the source term to be of ${\rm O}(p^2)$. Consequently, \pref{LexttensorChPT} has chiral dimension 2 as indicated by the superscript. In the mesonic chiral lagrangian the source term starts to contribute at chiral dimension 4. Therefore, the leading tensor field proportional to $\epsilon^{abc}\partial_{\mu}\pi^b\partial_{\nu}\pi^c$ stemming from it can be ignored for our purposes. 

As already stated in \cite{Cata:2007ns}, the power counting for the tensor is not motivated by physical arguments. In contrast to the counting rules for the scalar and pseudoscalar densities there is no physical realization of the symmetry breaking by a tensor in the QCD lagrangian that can be invoked to motivate the power counting $t_{\mu\nu} \sim {\rm O}(p^2)$. Other choices are possible, and any choice will affect the way operators with a different number of tensor sources are organized in the chiral expansion \cite{Cata:2007ns}. Still, irrespective of any particular counting rule the $N\pi$ Lagrangian in \pref{LexttensorChPT} will still be of smaller chiral dimension than the mesonic part. The reason is simple: The two Lorentz indices of the tensor source can be contracted with $\sigma^{\mu\nu}$ in the $N\pi$ Lagrangian, while two covariant derivatives of the pion field are necessary in the mesonic lagrangian. The latter is therefore of chiral dimension 2 higher. Essentially the same argument has been given in Ref.\ \cite{Dorati:2007bk} where an external {\em symmetric} tensor field was coupled to the QCD lagrangian.   

\end{appendix}

\end{document}